\documentclass[12pt]{article}

\usepackage{amsmath,amssymb,amscd,subfigure}
\usepackage{url}
\usepackage{graphicx} 

\def\beq{\begin{equation}}
\def\eeq{\end{equation}}
\def\barr{\begin{eqnarray}}
\def\earr{\end{eqnarray}}

\newcommand{\pI}{\mathsf{p1}} 
\newcommand{\pg}{\mathsf{pg}}
\newcommand{\pgg}{\mathsf{pgg}}
\newcommand{\pII}{\mathsf{p2}}
\newcommand{\pM}{\mathsf{pm}}
\newcommand{\cm}{\mathsf{cm}}
\newcommand{\pmm}{\mathsf{pmm}}
\newcommand{\cmm}{\mathsf{cmm}}
\newcommand{\pmg}{\mathsf{pmg}}
\newcommand{\pIII}{\mathsf{p3}}
\newcommand{\pIIImI}{\mathsf{p3m1}}
\newcommand{\pIIIIm}{\mathsf{p31m}}
\newcommand{\pIV}{\mathsf{p4}}
\newcommand{\pIVm}{\mathsf{p4m}}
\newcommand{\pIVg}{\mathsf{p4g}}
\newcommand{\pVI}{\mathsf{p6}}
\newcommand{\pVIm}{\mathsf{p6m}}

\newcommand {\Zb} {\mathbb{Z}} 
\newcommand {\Rb} {\mathbb{R}} 
\newcommand {\Db} {\mathbb{D}} 

\begin{document}

\begin{titlepage}

\begin{center}
\hfill  \quad  \\
\vskip 0.5 cm
{\Large \bf Fermi sea topology and boundary geometry for free particles in one- and two-dimensional lattices}

\vspace{0.5cm}

Guillermo~R.~Zemba\\
\medskip
{\em Facultad de Ingenier\'ia y Ciencias Agrarias,  Pontificia Universidad Cat\'olica Argentina,}\\
{\em  Av. Alicia Moreau de Justo 1500,(C1107AAZ) Buenos Aires, Argentina}\\
\medskip
{\em  and}\\
\medskip
{\em Departamento de F\'{\i}sica Te\'orica de Interacciones Fundamentales y Sistemas Complejos, Laboratorio Tandar,}\\
{\em  Comisi\'on Nacional de Energ\'{\i}a At\'omica,} \\
{\em Av.Libertador 8250,(C1429BNP) Buenos Aires, Argentina}\\
\end{center}
\vspace{.3cm}
\begin{abstract}
\noindent
Free gasses of spinless fermions moving on a lattice-symmetric geometric background are considered. 
Their topological properties at zero temperature can be used to classify their Fermi seas and associated boundaries. 
The flat orbifolds ${\Rb}^{d}/\Gamma$, where $\Gamma$ is the crystallographic group of symmetry in $d$-dimensional momentum space, are used to accomplish 
this task.
Two topological classes exist for $d=1$: an interval, which is identified as a conductor, and a circumference, which corresponds to an insulator.
The number of topological classes increases to 17 for $d=2$: 8 have the topology of a disk, that are generally recognized as conductors, and 4 correspond to a 2-sphere, matching insulators. Both sets eventually contain a finite number of conical singularities and reflection corners at the boundaries. 
The remaining cases in the listing relate to conductors (annulus, M\"obius strip) and insulators (2-torus, real projective plane, Klein bottle). Examples that fall under this list are given, along with physical interpretations of the singularities.
It is anticipated that the findings of this classification will be robust under perturbative interactions due to its topological character.
\end{abstract}
\vskip 0.5 cm
\end{titlepage}
\pagenumbering{arabic}
\section{Introduction} 
The theoretical understanding of the properties of materials from their microscopic structure has advanced significantly during the past few decades. Band theory \cite{lanlifs,ashmer,simons}, which provides a powerful framework for comprehending and forecasting the electronic properties of solids by describing the potential energy levels of electrons moving in a lattice structure \cite{single,zunger}, is one of the primary tools for this advancement. Materials can be categorized as conductors, insulators, or semiconductors based on the band structures that emerge from these investigations. Furthermore, these techniques can describe and predict the modification of material properties by changes in their microscopic structure.   
Determining a material's Fermi sea and Fermi surface is one of the most important outcomes of band structure research. For the sake of clarity, the term "Fermi boundary" will be used rather than "Fermi surface" since lattices in one and two spatial dimensions will be taken into consideration in this work.
For a material at zero temperature, the Fermi boundary divides occupied (low energy) from unoccupied electron states in momentum space for a given energy function. The paradigm of a free fermion gas is extended by this definition to the situation of a gas traveling on a spatial lattice background with no additional structure beyond its geometry. 
Furthermore, important details about the material properties, such as its electrical behavior, are compactly encoded by the shape and topology of the Fermi sea and its boundary \cite{fermirev}. Experimental methods like as angle-resolved photoemission spectroscopy (ARPES) and the detection of the De Haas-van Alphen effect \cite{fermiexp} support the expected momentum space features of the Fermi sea.
However, band studies typically need a significant amount of processing power. In other words, a large amount of detailed information supplied as an input results in a large amount of detailed information as an output.
 
Nonetheless, in the long-distance, low-energy domain of a system in the thermodynamic limit (large number of non-relativistic particles at constant spatial density), effective field theories (EFT) offer alternative, primarily analytical methods to study the motion of fermions in crystal backgrounds \cite{polch,boze}. These techniques yield a corresponding amount of information since they require far less precise microscopic data.
The main characteristics are that the result is universal information that does not depend on minute specifics and simply needs basic computation tools. 
The symmetries of the microscopic lattice essentially supply the input information for free particles. The free hamiltonian that describes the particle dynamics, {\it i.e.}, having only kinetic energy, defines the Fermi sea and energy. 
This free theory-based starting point is used in accordance with the Wilson Renormalization Group concepts and the EFT approach \cite{polch}. 
Therefore, it should be possible to ascertain the universal characteristics of the Fermi seas and their boundaries just from the lattice's symmetry contents. Leaving aside their specific quantitative characteristics, these should be the topological and geometrical structures of the Fermi seas and boundaries. 
It is anticipated that the topological structures of the Fermi seas and their boundaries will remain invariant when small interactions between the particles and a potential background (such as a crystal of ions) are included.
In the limit in which the spatial density of the particles $n(d)= N/V(d)$ is constant, {\it i.e.}, $N \to \infty$, $V(d) \to \infty$, where $V(d)$ is the $d$-dimensional volume, the free fermion gas picture describes the propagation of $N$ non-relativistic, non-interacting fermion particles in $d$-dimensional Euclidean space, which is frequently thought to be confined to a large volume. 
This image can be expanded to include the same kind of particles propagating in the presence of a background lattice \cite{boze}, which is not an interacting background potential but rather a modification of the free geometry.
The accessible space to be occupied by free fermionic states is known as the momentum space geometry, and it is changed from ${\Rb}^{d}$ to the quotient space ${\Rb}^{d}/\Gamma$, where $\Gamma$ is the lattice's crystallographic group of symmetry in $d$-dimensional momentum space. An orbifold is the name given to this topological space (see {\it e.g.}, \cite{nilse,caramello}).
A warning about this statement is that only flat compact connected orbifolds (for example, in $d=2$ this means that their Euler characteristic is zero) should be taken into consideration because they should locally resemble ${\Rb}^{d}$. 
This is due to the possibility of orbifolds having point singularities that alter their spatial curvature.

The $1d$ and $2d$ instances are shown and discussed in the next two sections. Prior to the presentation of Concluding Remarks, the $3d$ situation is briefly described.

\section{The $1d$ case}

$N$ non-interacting particles moving on the space ${\Rb}$ with coordinate $x$ and macroscopic size $L$ describe a gas of non-relativistic spinless free fermions of mass $m$ in $d=1$. The kinetic energy $H = p^{2}/(2m)$ determines their hamiltonian, and the solutions to the Schr\"odinger equation are free plane wave functions $e^{ipx}$ ($\hbar = 1$) with quantized momenta $p=2\pi n/L$, where $n$ is an integer or half-integer depending on the selected boundary conditions at $x=0$ and $x=L$. $E(p)$ is the energy, and the dispersion relation is $E(p) = p^{2}/(2m)$.
The configuration obtained by filling all the lowest energy $N$ states in momentum space to form a segment with boundary points -$p_F$ and $p_F$, where $p_F=\pi n(1)$ is the Fermi momentum, is the Fermi sea at zero temperature in the thermodynamic limit $N \to \infty$, $L \to \infty$ with $n(1)=N/L$ constant. Consequently, two points make up the Fermi boundary of the Fermi sea, which is an interval in momentum space \cite{lanlifs}. Keep in mind that $\Rb$ is the provided geometry of the space of possible momenta.

A background spatial lattice might be added to the above picture, but this would just alter the geometrical characteristics of free propagation ({\it i.e.}, 
no interaction with a crystal is implied).
The sole kind of Bravais lattice in $d=1$ is made up of equally spaced points along the $\Rb$ line with a microscopic spacing of $a$. The spatial (direct) lattice is Fourier transformed to yield the reciprocal or momentum lattice, which is found to be isomorphic to it with a large scale spacing of $2\pi/a$.
The hoping hamiltonian, a periodic variant of the free hamiltonian, takes its place:
\beq
H = -t[ \cos ( pa) -1]\ ,
\label{hhop}
\eeq 
where the parameter $t$ is connected to the particle mass by $t=1/(ma^2)$ after taking into account the limit of small momentum of (\ref{hhop}). The reflection ($P$) $\Zb_2$ symmetry is incorporated via (\ref{hhop}): $p \to -p$.
Bloch wave functions $u(x) e^{ikx}$ with $u(x+2\pi/a) =u(x)$ \cite{lanlifs,ashmer} are now the solutions of the Schr\"odinger equation, and the dispersion relation becomes $E(p) = -t[ \cos (pa) -1]$.
The number of particles per unit cell $\nu = Z = N/N_c $ and the number of spatial cells $N_c = L/a$ can be conveniently defined so that $n(1)=\nu /a$.
As a result, the momentum space states' geometry is realized as a ring with radius $1/a$. These fall into analogous regions known as Brioullin zones (BZ) due to the $2\pi/a$ periodicity of momenta. 
The circumference $S^1$, which represents the ring, is now the geometry of the space of available momenta. Because of the $T$ symmetry, it is possible to imagine filling the lowest momentum states with free fermions beginning with the circumference angle parametrization $\theta = 0$ up to a value $\theta$ for $N/2$ particles with positive momentum and -$\theta$ for the remaining $N/2$ particles with negative momentum. Consequently, the Fermi boundary is made up of two isolated points at $\pm p_F$ with $2p_F = (2\pi/L)N=(2\pi/a)\nu$, and the Fermi sea geometry is the interval $[-p_F, p_F ]$.
Instead, the fundamental domain of momenta, or the subset of Fermi sea points identified by the $P$ symmetry, provides the topology of the Fermi sea. It is represented by the irreducible wedge of the $1d$ lattice, or the arc $p > 0$. It should be noted that the image above can be folded around the abscissa axis. Therefore, the interval $(0,p_F]$ represents the space of free independent momentum states (the zero momentum state should be omitted due to the uncertainty principle given the finite size of the system). 
The orbifolding of the $S^1 $ manifold by the $P$ ($\Zb_2$) symmetry, or the quotient space $S^1/\Zb_2$, is the name given to this folding. 
Finally, the Fermi sea's generic topology is that of the {\it interval}. 
Because there are accessible particle-hole low-lying excitations around each of the Fermi points, the physical systems can be classified as {\it conductors}. 
However, a unique configuration is reached when the momentum ring's filled states precisely cover it, {\it i.e.}, when twice the Fermi momentum equals the lattice spacing $2\pi/a$, indicating that both Fermi points strike the first BZ's boundaries: $p_F=\pi n(1) = \pi /a$, {\it i.e.}, $\nu =1$. The Fermi sea becomes a circumference with no Fermi boundary because of the ring topology, which causes both Fermi points to coalesce (be identified).
The systems are therefore classified as {\it insulators} (more accurately, they could be called {\it geometric insulators}) since there are no low-lying excitations available.
The Fermi sea's topology is now determined by the {\it circumference} $S^1$.

A purely mathematical approach may have produced the same outcomes. 
Two alternative space groups $\Gamma$ in $d=1$ are $\Zb$ and $\Db_\infty$ (see Table 1) \cite{nilse}.
These result from identifying the isometries of $\Rb$ that serve as the group's generators. 
The circumference and the interval are the two orbifolds (quotient spaces) $\Rb/\Gamma$.
\begin{table}[ht!]
\begin{center}
\begin{minipage}[t]{0.7\textwidth}
\begin{center}
\vspace{0.5cm}
\begin{tabular}{|c||c|c|c|}
\hline
$\Zb$ & $\Rb/\Zb$ & $S^1$ & circumference  \\
$\Db_\infty$ & $\Rb/\Db_\infty$ & $S^1/\Zb_2$ & interval\\
\hline
\end{tabular}
\caption{The first column lists space groups $\Gamma$ for $d=1$. The matching orbifold is specified in the second. Their topologies are shown by the third and fourth.}
\end{center}
\end{minipage}
\end{center}
\end{table}
It is important to note that the space group $\Db_\infty$ can be defined equally well in terms of two $\pi$-rotations: $\Db_\infty \simeq \Zb_2 \times \Zb_2$. 
As a result, the classification of $d=1$ momentum space orbifolds produces the same Fermi sea topologies as the earlier study based on the straightforward physical picture.

By promoting as a guiding principle the construction of all momentum orbifold spaces obtained from $\Rb^{d}$ and $\Gamma$, where $\Gamma$ denotes the crystallographic symmetry group of the lattice and defines the available spaces to be filled by free fermions, this mathematical analysis can be extended to higher dimensional lattices (specifically, to $d=2$ and $d=3$). 


\section{The $2d$ case}

$N$ non-interacting particles moving over the space ${\Rb}^{2}$ with coordinates $(x,y)$ describe a gas of non-relativistic spinless free fermions of mass $m$ in $d=2$. A macroscopic size of $L^2$ is taken into consideration for simplicity. 
$H = \left( p_x^{2}+p_y^{2} \right)/(2m)$ is their free hamiltonian. With a $2d$ square lattice of momenta $(p_x,p_y)=(2\pi n_x/L,2\pi n_y/L)$ with $n_x$, $n_y$ integers, the construction is similar to that of the $d=1$ example. 
Because of the hamiltonian's rotational invariance \cite{lanlifs}, the Fermi sea's geometry at zero temperature is a circle with radius $p_F$. The formula for it is $p_F = 2\sqrt{\pi n(2)}$, where $n(2) = N/L^2 =\nu/a^2$ using $L^2 = N_c a^2$ and $\nu = Z = N/N_c$. 
After excluding the zero momentum state at the origin in the polar coordinates implied by the rotational symmetry about the origin, a more thorough examination of the topology of the Fermi sea reveals that it is given by the annulus $A$, which is a disk $D^{2}$ with a puncture (small hole) at the origin. This manifold is flat and has a zero Euler characteristic, which is a curvature measure. 
Keep in mind that the equally flat $\Rb^{2}$ is the provided geometry of the space of potential momenta.

Free fermions on a spatial square lattice of side $a$ define a straightforward expansion of the $1d$ example. A square with side $2\pi/a$ represents the momentum space lattice. Given this geometry, the free hoping hamiltonian is as follows:
\beq
H = -t[ \cos ( p_x a) + \cos ( p_y a) -2]\ .
\label{hhop2}
\eeq 
Instead of rotational invariance in momentum space, this hamiltonian has the discrete symmetries $p_x \leftrightarrow p_y$, $p_x \leftrightarrow -p_x$, and $p_y \leftrightarrow -p_y$. In momentum space, the Fermi sea is shaped like a rounded diamond for generic $\nu < 1$ and like a proper diamond for $\nu=1/2$.
The square of the side intervals $(-\pi/a,\pi/a)$ in the $p_x$ and $p_y$ axes yields the first BZ. The Fermi momentum $p_F$ for $\nu=1/2$ is equal to half the diamond's side length. It features four corners at an angle of $\pi /2 $ at $(\pm \pi /a, \pm \pi /a)$. These corners can be thought of as Van Hove singularities, where the Fermi velocity (the gradient of the energy with respect to the momentum) disappears \cite{polch}. 
Therefore, when $\nu < 1$, the Fermi sea's geometry is that of a sharp or rounded diamond, which generally characterizes a {\it conductor}. 
There is a unique configuration for $\nu = 1$, just like in the $d=1$ case. With the diamond's vertices at $\sqrt{2} (-\pi/s,\pi/a)$ and $p_F = \pi/a$, the Fermi sea extends past the first BZ. By adding a vector corresponding to $(\pm 2\pi/a, \pm 2\pi/a)$, the opposing sides of the Fermi sea can be mapped into one another, indicating that they are equivalent in the lattice and, therefore, identifiable (the so-called {\it nesting} method).The Fermi sea takes on the topology of a $2d$ torus $T^2$, which is thought to be an {\it insulator} and has no boundaries. There are numerous examples that could be specifically taken into account in $2d$.

However, the mathematical analysis is more potent and proceeds as follows.  
$2d$ Bravais lattices come in five different varieties: oblique (no constraints on the angles or lengths of the unit cell vectors), rectangular ($\pi/2$ angles, unequal side lengths), centered rectangular ($\pi/2$ angles, unequal side lengths, centered unit cell), square ($\pi/2$ angles, equal side lengths), and hexagonal ($2\pi/3$ and $\pi/3$ angles, equal side lengths) \cite{lanlifs,ashmer}.Each $2d$ Bravais lattice's reciprocal is also a $2d$ Bravais lattice with the same symmetry as the direct one. 
The lattice's (rotational, reflectional, etc.) symmetries are described by the point groups. 
There are 17 different crystallographic space groups (sometimes called wallpaper groups) in $2d$. 
Bravais lattices with various point group symmetries are combined to create these groups. Translations, reflections, $2 \pi / n$-rotations of order $n=2,3,4,6$, and so-called glide-reflections—translations with a simultaneous mirror reflection—make up the set of isometries. 
While Conway's orbifold notation~\cite{Conway} is more commonly employed in mathematics, the crystallographic notation~\cite{CRC} is frequently utilized in physics literature for the space groups.
Each Bravais lattice type and its potential point group symmetries are taken into consideration while classifying the 17 wallpaper groups \cite{sasse}. The list includes:
\noindent
\begin{enumerate}
\item 15 primitive lattices:
\begin{enumerate}
        \item 2 parallelogram lattices: $\pI, \pII$
        \item 5 rectangular lattices: $\pM, \pg, \pmm, \pmg, \pgg$
        \item 3 square lattices: $\pIV, \pIVg, \pIVm$
        \item 3 trigonal lattices: $\pIII, \pIIIIm, \pIIImI$
        \item 2 hexagonal lattices: $\pVI, \pVIm$
\end{enumerate}
\noindent
\item 2 centered lattices: $\cm,\cmm$
\end{enumerate}
The discrete group names in the crystallographic conventions (Hermann-Mauguin symbols) consist of a maximum of four letters and numbers, such as $\pVIm$.
The first is either $\mathsf{o}$ (centered cell, with an extra lattice point in its center) or $\mathsf{p}$, which stands for primitive cell. With the previously mentioned conventions, the second represents the highest order rotation. The third character, $\mathsf{m}, \mathsf{1}, \mathsf{g}$, represents reflection, neither of both, or glide reflection (reflection followed by a translation) with regard to the main axis. 
The fourth is comparable to the third, but it refers to the secondary axis and once more accepts the values $\mathsf{m}, \mathsf{1}, \mathsf{g}$.
Reflections or glide reflections are absent if the third or fourth characters are missing \cite{CRC,sasse}. Some
symmetry group isomorphisms are: $\pI \simeq  \Zb^2$, $\pM \simeq  \Zb \times \Db_\infty$ and 
$\pmm \simeq \Db_\infty  \times \Db_\infty$.
Conway's orbifold notation~\cite{Conway}, on the other hand, uses a finite string of characters to indicate a group: an integer $\mathsf{n}$ to the left of an asterisk $\mathsf{*}$ indicates a rotation of order $n$, the asterisk indicates a reflection; an integer $\mathsf{n}$ to the right of the asterisk indicates a rotation of order $2n$ and reflects through a line; and $\mathsf{\times}$ indicates a glide reflection. 

Consequently, the quotient spaces $\Rb^2/\Gamma$, where $\Gamma$ is one of the space groups given in Table 2 \cite{nilse,lim}, yield 17 (compact and linked) orbifolds in $2d$. \begin{table}[h!]
\centering
\begin{center}
\begin{minipage}[t]{0.98\textwidth}
\begin{center}
\begin{tabular}{|c|c||c|c|c|c|c|}
\hline
$\pII$ & $2222$ & $\Rb^2/\pII$ & 4-pillow & $S^{2}$ & 2,2,2,2 & \\
$\pIII$ & $333$ & $\Rb^2/\pIII$ &  3-pillow & & 3,3,3 & \\
$\pIV$ & $442$ & $\Rb^2/\pIV$ &  3-pillow & & 2,4,4 & \\
$\pVI$ & $632$ & $\Rb^2/\pVI$ &  3-pillow & & 2,3,6 & \\
\hline
$\pmm$ & $*2222$ & $\Rb^2/\Db_\infty^2$ & rectangle & $D^{2}$ &  & 2,2,2,2\\
$\cmm$ & $2*22$ & $\Rb^2/\cmm$ & triangle & & 2 & 2,2 \\
$\pmg$ & $22*$ & $\Rb^2/\pmg$ & open 4-pillow & & 2,2 & 2\\
$\pIIImI$ & $*333$ & $\Rb^2/\pIIImI$ &  triangle & & & 3,3,3\\
$\pIIIIm$ & $3*3$ & $\Rb^2/\pIIIIm$ &  open 3-pillow &  & 3 & 3\\
$\pIVm$ & $*442$ & $\Rb^2/\pIVm$ &  triangle & & & 2,4,4\\
$\pIVg$ & $4*2$ & $\Rb^2/\pIVg$ & open 3-pillow & & 4 & 2\\
$\pVIm$ & $*632$ & $\Rb^2/\pVIm$ &  triangle & & & 2,3,6\\
\hline
$\pI$ & $\circ$ & $\Rb^2/\Zb^2$ & torus & $T^2$ & & \\
$\pg$ & $\times \times$ & $\Rb^2/\pg$ &  Klein bottle & $K$ & & \\
\hline
$\pgg$ & $22 \times$ & $\Rb^2/\pgg$ &  real projective plane & $\Rb P^2$ & 2,2 &  \\
\hline
$\pM$ & $**$ & $\Rb^2/(\Zb \times \Db_\infty)$ &annulus & $A$ & &  \\
$\cm$ & $* \times $ & $\Rb^2/\cm$ & M\"obius strip & $M$ & &  \\
\hline
\end{tabular}
\centering
\caption{The $2d$ space groups $\Gamma$ are listed in both orbifold~\cite{Conway} and crystallographic~\cite{CRC} notation in the first two columns. The matching quotient spaces $\Rb^2/\Gamma$ are shown in the next column. The orbifold spaces' general topology is seen in the fourth and fifth columns.
The order of conical points and the order of corner reflections are listed in the sixth and seventh, respectively.}
\end{center}
\end{minipage}
\end{center}
\end{table}
The topological classes of the Fermi seas and their boundaries are shown in the final three columns of Table 2.
Each orbifold described in Table 2 is shown schematically on pages 79–80 of \cite{dunbar}.
The 2-sphere ($S^2$), the 2-disk ($D^2$), the real projective plane $\Rb P^2$, the 2-torus ($T^2$), the Klein bottle ($K$), the annulus ($A$), and the M\"obius strip ($M$) are the seven fundamental topologies. These are all Euclidean planar spaces, meaning that they have zero Euler characteristics.  
In the latter four classes, this feature is automatic, but not in the initial three. 
In fact, a few conical singularities of order 2 are required for the real projective plane to become flat.
Additionally, in order for the final space to be flat, the spaces $S^2$ and $D^2$ also need a certain number of isolated singularities; this leads to 4 $S^2$ and 8 $D^2$ subclasses.
These possibilities are determined from the Riemann-Hurwitz
formula of the Euler characteristic for an orbifold $O$, which is given by (see, {\it e.g.}, \cite{lim}): 
\beq
\chi(O) = \chi(M) - \sum_{i=1}^{p} \left( 1- \frac{1}{m_i} \right) - \frac{1}{2}\sum_{j=1}^{q} \left( 1- \frac{1}{n_j} \right)
\label{euler}
\eeq
where $\chi(M)$ is the Euler characteristic of the smooth base manifold, $p$ are the cone points of order 
$m_1 , m_2, \dots , m_p $ and $q$ the corner reflectors of order $n_1 , n_2, \dots , n_q $. 

Seven orbifolds are identified generically as {\it insulators} (no Fermi boundary) on the listing in Table 2: the four mentioned as $S^{2}$, the torus $T^{2}$, the real projective plane, and the Klein bottle (the final two are not orientable in $3d$). The remaining ten are identified generically as {\it conductors} (with a Fermi boundary): the annulus, the M\"obius strip (non-orientable in $3d$), and the eight listed as $D^{2}$.
Van Hove singularities can be recognized as reflection singularities at the Fermi boundary \cite{ashmer,polch}.
There may be some selective, fine-tuned conduction spots with fixed Fermi velocities if the conical singularities found in the bulk of the Fermi sea are identified as Dirac points \cite{conical}. 
Furthermore, materials such as twisted graphene with transition metal dichalcogenide (TMD) layers have been explored for the emergence of  M\"obius-like Fermi surfaces \cite{moebius}.
As a consistency test, it is observed that this classification includes the diamond-shaped and 2-torus Fermi seas found in the square lattice example. The natural extensions of the $1d$ case, with symmetries $\Rb/\Db_\infty$ (interval) and $\Rb/\Zb$ (circumference), correspond to the group $\pmm$ with symmetry $\Rb^2/\Db_\infty^2$ in the first instance and the group $\pI$ with symmetry $\Rb^2/\Zb^2$ in the second.

The process of using the mathematical classification of $2d$ flat orbifolds determines the corresponding $2d$ lattice and its symmetry from which they arise, in addition to the list of potential Fermi sea topologies and boundary geometries. 
However, a more straightforward argument might be used to independently derive the former information. 
It should be noted that all of the many subcases of the first two topological cases in Table 2—the disk with genus 1 and the 2-sphere with genus 2—may have been discovered by listing every potential singularity on these fundamental topologies after the flatness constraint was imposed. 
Since there are no boundaries in the case of $S^{2}$, there are no potential Van Hove singularities. Only conical ones of orders 2, 3, 4, and 6 are feasible. Since $S^{2}$ has an Euler characteristic of 2, the singularities should reduce this number to 0 and so add -2 to (\ref{euler}). After using (\ref{euler}), the only feasible combinations that match the contents of Table 2 are (3,3,3), (2,4,4), (2,3,6), and (2,2,2,2). 
Since the boundary is now the closed curve $S^{1}$, there are more options for the situation of $D^{2}$ with the Euler characteristic 1.  
Three triangles of internal angles $(\pi/3,\pi/3,\pi/3)$, $(\pi/2,\pi/4,\pi/4)$ and $(\pi/2,\pi/3,\pi/6)$, and one diamond of internal angles $(\pi/2,\pi/2,\pi/2,\pi/2)$ represent the four options that do not contain bulk singularities.
The total contribution of all these configurations to the Euler characteristic is -1. Additionally, a droplet (tear) with a boundary angle of $\pi/2$ and a bulk singularity of order 4, another similar droplet with a corresponding boundary angle of $\pi/3$ and a bulk singularity of order 3, and a droplet with two boundary angles of $\pi/2$ and a bulk singularity of order 2 are the three possibilities that combine boundary and bulk singularities. Lastly, a configuration with just two order 2 bulk singularities.

Note that \cite{catast} provides a related categorization of Fermi surface topological transitions in $2d$ based on Catastrophe theory. Table 2's contents can be used to explain these transitions as topological operations connecting list pairs. For example, by combining the two conical singularities of order 2 into one of order 1 (that is, a puncture or another boundary with $S^{1}$ topology), the $D^{2}$ configuration with two bulk singularities of order 2 might be transformed into that of an annulus.  

\section{The $3d$ case: outline}

There are fourteen distinct Bravais lattices in $d=3$. Seven crystal systems with various centering kinds are combined to create these lattices. There are 230 space groups. It is possible to do an analysis akin to that of the $d=1$ and $d=2$ situations, but it has been put off for future study.

\section{Concluding remarks}

Supported by a flatness condition, the classification of the topology of Fermi seas by taking into account the orbifolds ${\Rb}^{d}/\Gamma$, where $\Gamma$ is the crystallography group of symmetry in $d$-dimensional momentum space, results in a listing that satisfies multiple consistency conditions and could be used as an ordering principle for future research on this subject. 

By imposing the flatness of the resulting orbifold, a simple enumeration of all the possible conical and reflection singularities on surfaces with positive curvature (of genus $g=2,1$, that is, the 2-sphere and the disk) can yield a list of potential topologies of the Fermi seas and the geometry of their boundaries in $d=2$. 

%
\def\NP{{\it Nucl. Phys.\ }}
\def\PRL{{\it Phys. Rev. Lett.\ }}
\def\PL{{\it Phys. Lett.\ }}
\def\PR{{\it Phys. Rev.\ }}
\def\IJMP{{\it Int. J. Mod. Phys.\ }}
\def\MPL{{\it Mod. Phys. Lett.\ }}

\end{document}